\documentclass[reprint,superscriptaddress,amsmath,amssymb,aps]{revtex4-2}

\usepackage{graphicx}
\usepackage[caption=false]{subfig}
\usepackage{multirow}
\usepackage{dcolumn}
\usepackage{bm}
\usepackage{hyperref}
\hypersetup{colorlinks=True,
            linkcolor=blue,
           anchorcolor = blue,
            citecolor = blue,
            filecolor = blue,
            urlcolor = blue }       

\begin{document}
\title{Unconventional photon blockade in cavity QED with parametric amplification }
\author{Madan Mohan Mahana}
\affiliation{Department of Physics, Indian Institute of Technology Guwahati, Guwahati 781039, Assam, India}
\author{Sanket Das}
\affiliation{Quantum Machines Unit, Okinawa Institute of Science and Technology Graduate University, Okinawa 904-0495, Japan}
\author{Tarak Nath Dey}
\email{tarak.dey@iitg.ac.in}
\affiliation{Department of Physics, Indian Institute of Technology Guwahati, Guwahati 781039, Assam, India}

\begin{abstract}
      We theoretically investigate the quantum-interference-induced photon blockade effect in a single two-level atom-cavity quantum electrodynamics (QED) system with degenerate parametric amplification. The analytical calculations reveal the optimal parametric gain and phase parameters for achieving optimum unconventional photon blockade conditions. Under the optimal parameter regime, the numerical results of the second-order correlation function demonstrate strong photon antibunching consistent with the analytical results. Furthermore, the numerical results corroborate that coherently driving the atom leads to a stronger photon blockade than a coherently driven cavity with the optimal parameters. We numerically demonstrate that the UPB effect is compromised by a non-zero cavity-atom coupling in the cavity-driven configuration. However, stronger photon antibunching can be attained with a non-zero cavity-atom coupling in the atom-driven configuration. This work may be suitable for experimentally realising a strongly antibunched single-photon source for applications in quantum technology.
\end{abstract}
\maketitle

\section{\label{sec:level1}INTRODUCTION}
 \noindent Photon blockade is a fundamental phenomenon in quantum optics that inhibits multiple-photon transmission. The blockade arises from photon-photon interaction in a nonlinear optical medium \cite{PhysRevLett.79.1467}. In a two-level atom-cavity system, a significant nonlinearity emerges in a strong coupling regime and exhibits anharmonicity in the discrete energy levels. A single photon with a specific energy excites the system from the ground to the first excited state. However, a second photon with identical energy can not be absorbed as it becomes off-resonant with the energy gap between the first and second excited states, commonly known as the conventional photon blockade (CPB) \cite{PhysRevA.102.033713, Lin:20, PhysRevLett.106.243601}. The CPB effect is associated with nonclassical photon statistics, and antibunching \cite{RevModPhys.54.1061, https://doi.org/10.1002/lpor.201900279,PhysRevA.103.053710,PhysRevLett.88.023601}. Recent experimental advances demonstrate the realizations of CPB in single atom-cavity QED systems \cite{Birnbaum2005,PhysRevLett.118.133604,Tang:21}, superconducting quantum circuits \cite{PhysRevLett.106.243601}, and single quantum dot-microcavity systems \cite{PhysRevLett.98.117402, Hennessy2007}.

Although systems relying on CPB are steadily improving, they pose significant technological challenges regarding scalability \cite{PhysRevA.78.062336}. On the other hand, weakly nonlinear systems occur more naturally and can be advantageous. Recent research realizes photon blockade in systems characterized by weak nonlinearity, originating from the medium itself or through a weakly coupled nonlinear medium \cite{PhysRevA.96.053810, Liu:20, PhysRevA.92.023838, PhysRevA.100.053832}. This phenomenon is named unconventional photon blockade (UPB). It occurs due to destructive interference between various excitation pathways that an optical system can follow to absorb or emit photons. This concept has recently been explored in several hybrid quantum systems, including cavity optomechanics \cite{Xu_2013, PhysRevA.92.033806, PhysRevA.98.013826}, cavity magnoics \cite{PhysRevA.94.063853, PhysRevA.93.063861, PhysRevA.101.042331, PhysRevA.103.052411, PhysRevA.101.063838}, superconducting quantum circuits \cite{PhysRevLett.121.043602}, and quantum dot cavity systems \cite{PhysRevLett.121.043601}. In cavity optomechanics and cavity magnoics, coupling cavity photons and either mechanical motion or the quantized magnetization modes of the ferrimagnetic yttrium iron garnet (YIG) sphere establishes excitation pathways for photonic excitations. Conversely, the interaction between microwave photons and nonlinear superconducting elements, such as Josephson junctions, generates multiple photonic excitation pathways in superconducting quantum circuits. In quantum dot cavity systems, the Jaynes-Cummings interaction between optical photons and semiconductor quantum dots or artificial atoms leads to these photonic excitation pathways.

Photon blockade can be used to engineer single-photon devices such as photon-turnstiles \cite{doi:10.1126/science.290.5500.2282}, single-photon transistors \cite{doi:10.1126/science.aat3581}, and quantum gates \cite{Hacker2016}. A harmonic cavity weakly coupled to a two-level atom is a vital setup to study the UPB effect \cite{Tang2015, PhysRevA.100.063817, PhysRevA.100.063834}.  One can enhance the UPB effect in such a system with parametric amplification \cite{PhysRevA.106.023704,mi14112123}. The global phase in a system plays a vital role in controlling quantum-interference-induced optical phenomena such as UPB, electromagnetically induced transparency (EIT), etc. \cite{Tang2015, PhysRevLett.105.073601, PhysRevA.91.043843, PhysRevA.103.052411}. The parametric gain and relative phase of the parametric pump field with respect to a weak coherent drive, applied to the two-level atom or cavity, play a crucial role in controlling the UPB effect in such a system \cite{PhysRevA.96.053827,Wang:20, PhysRevA.102.043705, PhysRevA.109.043712}.  

This paper investigates the UPB effect in a cavity containing a two-level atom. We employ a weak, coherent light source to drive the cavity or the atom. In addition to that, we pump the cavity by a degenerate parametric amplifier (DPA). Our analytical calculations dictate how the UPB effect with strong photon antibunching can be manifested by tuning the parametric gain and phase parameters to the optimal values. Our study highlights the advantage of driving the atom over driving the cavity for stronger photon antibunching under optimal UPB conditions. We also demonstrate numerically that, in the cavity-driven configuration, a non-zero cavity-atom coupling diminishes the UPB effect. In contrast, the atom-driven configuration benefits from a non-zero cavity-atom coupling, enabling enhanced photon antibunching.
 
The paper is organized as follows. In section. \ref{sec:level2}, a theoretical model of a single two-level atom-cavity system is described. In section. \ref{sec:level3}, we
employ the amplitude method to obtain the optimal conditions for UPB for both the cavity and atom-driven cases. Section \ref{sec:level4} numerically validates our analytical findings. Finally, we summarize the key findings of the work in section \ref{sec:level5}.
\section{\label{sec:level2}Theoretical Model}
\begin{figure}[t!]
    \centering
    \includegraphics[width=0.4\textwidth]{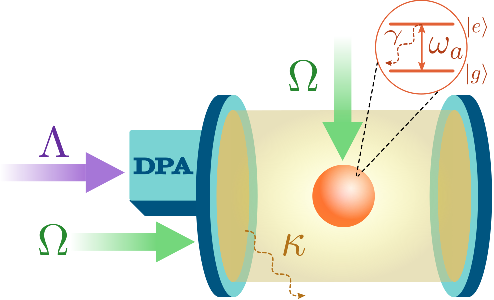}
    \caption{(Color online) Schematic diagram of a single two-level atom cavity QED system, parametrically pumped via a degenerate parametric amplifier. The cavity or the atom is driven by a coherent light source with Rabi frequency $\Omega$.}
    \label{fig:fig1}
\end{figure}

\noindent We consider a single two-level atom-cavity QED system, where the cavity is parametrically pumped via a DPA as proposed in \cite{PhysRevA.106.023704,mi14112123}. A coherent pump field drives the cavity or the two-level atom. The proposed model system is shown schematically in Fig. \ref{fig:fig1} that can be experimentally realized with an optical cavity containing a single atom pumped by an Optical parametric oscillator  \cite{PhysRevA.58.4056} or a superconducting qubit-resonator system pumped by a Josephson parametric amplifier \cite{Murch2013, PhysRevX.6.031004}. In the rotating frame with frequency $\omega_d$, the whole system can be described by the Hamiltonian
\begin{align}\label{1}
  H=&\hbar\Delta_c a^\dag a+ \hbar\Delta_a\sigma^\dagger \sigma+\hbar\chi (a^\dag\sigma+a\sigma^\dagger)\nonumber\\
  &+\hbar\Lambda({a^\dagger}^2 e^{-i\phi}+a^2 e^{i\phi})+H_d,
\end{align}
where $\Delta_c=(\omega_c-\omega_d)$ and $\Delta_a=(\omega_a-\omega_d)$ denote the frequency detunings between the cavity and two-level atom with driving fields, respectively. All external driving fields' frequencies are considered the same ($\omega_d$) for brevity. The annihilation and creation operators of the cavity are denoted by $a$ and $a^\dagger$. Similarly, the atomic lowering and raising operators are $\sigma=|g\rangle\langle e|$ and $\sigma^\dagger=|e\rangle\langle g$. The states $|e\rangle$ and $|g\rangle$ denote the excited state and the ground state of the two-level atom, respectively. The first two terms of the Hamiltonian represent the free energies of the cavity and the atom, respectively. The cavity interacts with the atom with a coupling strength $\chi$ as given in the third term. Depending on whether a coherent field pumps the cavity or the atom, we can write $H_d=\hbar\Omega(a^\dagger+a)$ or $H_d=\hbar\Omega(\sigma^\dagger+\sigma)$, where $\Omega$ is the Rabi frequency of the coherent pump field.  The parametric gain of the DPA pumping the cavity is denoted by $\Lambda$, and $\phi$ is the phase difference between the parametric pump and the coherent drive applied to the cavity or the atom. The dynamics of the system can be analyzed by solving the Lindblad master equations for the density matrix $\rho$ \cite{RevModPhys.82.1155}, given by
\begin{align}\label{1a}
 \dot{\rho}=\frac{1}{i\hbar}[H,\rho]+\kappa\mathcal{L}(a)\rho+ \gamma\mathcal{L}(\mathcal{\sigma})\rho,
\end{align}
where $\mathcal{L}(\mathcal{O})\rho = (2\mathcal{O}\rho\mathcal{O}^\dagger-\rho\mathcal{O}^\dagger\mathcal{O}-\mathcal{O}^\dagger\mathcal{O}\rho)/2$ is the Lindblad superoperator which takes into account the effect of decoherence in the system dynamics. The decay rates of the cavity and the two-level atom are denoted by $\kappa$ and $\gamma$, respectively. In the following section, we derive the analytical results to determine the optimal conditions for UPB in the proposed model system.
\section{\label{sec:level3}Analytical results}
\noindent The UPB effect refers to the suppression of multiphoton excitation due to quantum interference between various excitation pathways \cite{PhysRevA.96.053810}. Single-photon blockade is associated with strong photon antibunching, which is characterized by the equal time second-order quantum correlation function $\mathrm{g}^{(2)}(0)=\langle a^\dagger a^\dagger a a\rangle/\langle a^\dagger a\rangle^2$ in the steady state. The value of $\mathrm{g}^{(2)}(0)$ is proportional to the probability of detecting two photons at the same instant of time. For a coherent source of light $\mathrm{g}^{(2)}(0)=1$, the super-Poissonian bunched light and the sub-Poissonian antibunched (non-classical) light correspond to $\mathrm{g}^{(2)}(0)>1$, and $\mathrm{g}^{(2)}(0)<1$, respectively. Photon blockade is associated with strong photon antibunching characterized by $\mathrm{g}^{(2)}(0)\ll 1$. Photon blockade can be realized in the proposed model system when the two-photon excitation pathways destructively interfere such that the second excited state of the cavity field remains unpopulated in the steady state. We analytically derive the optimal conditions for photon blockade in the proposed model system for the cavity-driven and atom-driven cases described by the Hamiltonian in Eq. \ref{1}.
\subsection{\label{sec:level3a}Cavity-driven case}
\begin{figure}[t!]
    \centering
    \includegraphics[width=0.45\textwidth]{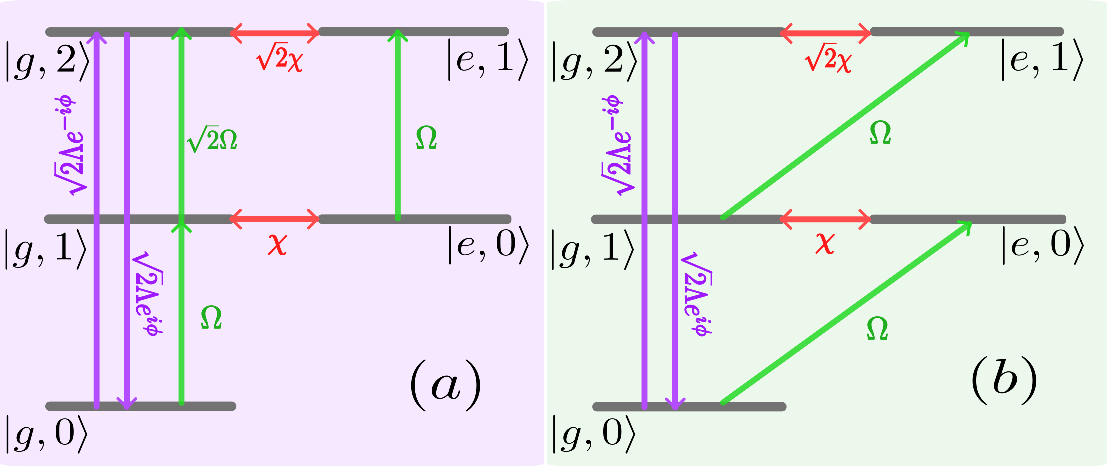}
    \caption{(Color online) Transition pathways for (a) the cavity-driven system and (b) the atom-driven system in the single-photon subspace.}
    \label{fig:fig2}
\end{figure}

\noindent Let us consider the cavity is coherently driven by a classical pump field in a two-level atom-cavity QED system with parametric amplification. The driving Hamiltonian $H_d=\hbar\Omega(a^\dagger+a)$ can be added to the Hamiltonian in Eq. \ref{1} to describe the system. The possible transition pathways for the cavity-driven case are shown in Fig. \ref{fig:fig2} (a). One can truncate the Hilbert space to only a two-photon excitation subspace for weak driving conditions $\Omega,\Lambda\ll \kappa$ with weak cavity-atom coupling $\chi\sim\kappa$. Following the amplitude method \cite{PhysRevA.83.021802}, we consider the wave-function ansatz
\begin{align}\label{2}
 |\psi\rangle=\sum\limits_{n=0}^2 C_{g,n}|g,n\rangle+\sum\limits_{n=0}^1 C_{e,n}|e,n\rangle, 
\end{align}
where $|C_{\alpha,n}|^2$ refers to the probability of detecting the state $|\alpha,n\rangle$ for $\alpha=e,g$ and $n=0,1,2$. Following the amplitude method, the dynamical equations for the probability amplitudes can be obtained by solving the Schr\"{o}dinger equation 
\begin{align}\label{3}
 i\hbar\frac{\partial|\psi\rangle}{\partial t}=H_1|\psi\rangle, 
\end{align}
where $H_1$ is the effective Hamiltonian 
\begin{align}\label{4}
 H_1/\hbar=H/\hbar-\frac{i\kappa}{2}a^\dagger a-\frac{i\gamma}{2}\sigma^\dagger \sigma.
\end{align}
The second and third terms in Eq. \ref{4} are phenomenologically added to consider the effect of decoherence. Substituting Eq. \ref{4} in Eq. \ref{3}, the dynamical equations for the probability amplitudes can be obtained as
\begin{align}\label{5}
 \dot{C}_{g,1} &= -i\Delta_c^\prime C_{g,1}-i\chi C_{e,0}-i\Omega C_{g,0}-\sqrt{2}i\Omega C_{g,2}\nonumber\\
 \dot{C}_{e,0} &= -i\Delta_a^\prime C_{e,0}-i\chi C_{g,1}-i\Omega C_{e,1}\nonumber\\
 \dot{C}_{e,1} &= -i(\Delta_c^\prime+\Delta_a^\prime) C_{e,1}-\sqrt{2}i\chi C_{g,2}-i\Omega C_{e,0}\nonumber\\
 \frac{\dot{C}_{g,2}}{\sqrt{2}} &= -\sqrt{2}i\Delta_c^\prime C_{g,2}-i\chi C_{e,1}-i\Omega C_{g,1}-i\Lambda e^{-i\phi}C_{g,0}
\end{align}
where $\Delta_c^\prime=\Delta_c-i\kappa/2$, and $\Delta_a^\prime=\Delta_a-i\gamma/2$. Under weak driving conditions, with perturbative approximations $|C_{g,0}|\gg|C_{g,1}|,|C_{e,0}|\gg|C_{g,2}|,|C_{e,1}|$, and $|C_{g,0}|\approx 1$, the steady-state solutions of the probability amplitudes can be obtained. Since we are interested in the second-order quantum correlation function

\begin{align}\label{6}
 \mathrm{g}^{(2)}(0)&=\frac{\langle \psi|a^\dagger a^\dagger a a|\psi\rangle}{\langle\psi| a^\dagger a|\psi\rangle^2}\approx\frac{2|C_{g,2}|^2}{|C_{g,1}|^4},
\end{align}
we give only the steady state probability amplitudes for the states $|g,2\rangle$  and $|g,1\rangle$ as
\begin{align}
 C_{g,2}&=\frac{\Lambda e^{-i\phi}(\Delta_a^\prime+\Delta_c^\prime)(\chi^2-\Delta_c^\prime\Delta_a^\prime)+M\Omega^2}{\sqrt{2}(\chi^2-\Delta_c^\prime\Delta_a^\prime)(\chi^2-\Delta_c^\prime(\Delta_a^\prime+\Delta_c^\prime)}\label{7a}\\
 C_{g,1}&=\frac{\Delta_a^\prime\Omega}{(\chi^2-\Delta_c^\prime\Delta_a^\prime)}\label{7b}
\end{align}
where $M=(\chi^2+\Delta_a^\prime(\Delta_a^\prime+\Delta_c^\prime))$. The optimal photon blockade can be achieved when $\mathrm{g}^{(2)}(0)\rightarrow0$ which implies $|C_{g,2}|=0$. Therefore, equating the numerator of Eq. \ref{7a} to zero can give us the optimal conditions for UPB. The optimal values of the parameters $\Lambda$ and $\phi$ can be obtained as 
\begin{align}\label{8}
\Lambda_c^{opt} e^{-i\phi_c^{opt}}=\frac{(\chi^2+\Delta_a^\prime(\Delta_a^\prime+\Delta_c^\prime))\Omega^2}{(\Delta_a^\prime+\Delta_c^\prime)(\Delta_c^\prime\Delta_a^\prime-\chi^2)},
\end{align}
where the parameters $\Lambda_c^{opt}$ and $\phi_c^{opt}$ are the modulus and the argument of the complex quantity in the right-hand side of Eq. \ref{8} for the cavity-driven case.

\subsection{\label{sec:level3b}Atom-driven case}

\noindent This section discusses the significance of parametric cavity drive in the atom-driven scenario. To our knowledge, the importance of parametric drive in this context has not been investigated till now \cite{PhysRevA.106.023704,mi14112123}. Following a similar procedure used for the cavity-driven case, one can obtain the optimal parameters for the atom-driven case. First, we substitute the atomic drive $H_d=\hbar\Omega(\sigma^\dagger+\sigma)$ in Eq. \ref{1}. Fig. \ref{fig:fig2} (b) illustrates the possible transition pathways for the atom-driven case. Following the amplitude method, we obtain the dynamical equations for the probability amplitudes 
\begin{align}\label{9}
 \dot{C}_{g,1} &= -i\Delta_c^\prime C_{g,1}-i\chi C_{e,0}-i\Omega C_{e,1} \nonumber\\
 \dot{C}_{e,0} &= -i\Delta_a^\prime C_{e,0}-i\chi C_{g,1}-i\Omega C_{g,0} \nonumber\\
 \dot{C}_{e,1} &= -i(\Delta_c^\prime+\Delta_a^\prime) C_{e,1}-\sqrt{2}i\chi C_{g,2}-i\Omega C_{g,1}\nonumber\\
 \dot{C}_{g,2} &= -2i\Delta_c^\prime C_{g,2}-\sqrt{2}i\chi C_{e,1}-\sqrt{2}i\Lambda e^{-i\phi}C_{g,0}
\end{align}
Solving these dynamical equations with perturbative approximations in the steady-state limit, we get
\begin{align}
 C_{g,2}&=\frac{\Lambda e^{-i\phi}(\Delta_a^\prime+\Delta_c^\prime)(\chi^2-\Delta_c^\prime\Delta_a^\prime)+\chi^2\Omega^2}{\sqrt{2}(\chi^2-\Delta_c^\prime\Delta_a^\prime)(\chi^2-\Delta_c^\prime(\Delta_a^\prime+\Delta_c^\prime)}\label{10a}\\
 C_{g,1}&=-\frac{\chi\Omega}{(\chi^2-\Delta_c^\prime\Delta_a^\prime)}\label{10b}
\end{align}
Now, the optimal condition for UPB can be expressed as
\begin{align}\label{11}
\Lambda_a^{opt} e^{-i\phi_a^{opt}}=\frac{\chi^2\Omega^2}{(\Delta_a^\prime+\Delta_c^\prime)(\Delta_c^\prime\Delta_a^\prime-\chi^2)},
\end{align}
where $\Lambda_a^{opt}$ and $\phi_a^{opt}$ denote the optimal gain and relative phase difference of the parametric pump field for the UPB effect in the atom-driven case. 

The optimal parameters for UPB in the cavity-driven and atom-driven cases given in Eq. \ref{8} and Eq. \ref{11}, respectively, are among the main results of this paper. The optimal parametric gains $\Lambda_c^{opt}$ ($\Lambda_a^{opt}$) are the same and the optimal parametric phases $\phi_c^{opt}$ ($\phi_a^{opt}$) are different for equal red and blue detunings in the cavity-driven (atom-driven) case. The optimal parameters depend on the cavity-atom coupling strength $\chi$, the drive-detuning $\Delta$, and the decay rates of the cavity and the atom. For simplicity, we consider $\Delta_c=\Delta_a=\Delta$ in the rest of the paper. In the following section, we study the effect of these parameters on the optimal UPB conditions.

\section{\label{sec:level4}Numerical results}

\noindent In this section, we present a comprehensive analysis of the UPB effect by numerically solving the Lindblad master equation Eq. \ref{1a}. To understand the strong photon antibunching effect, we numerically calculate the second-order quantum correlation function $\mathrm{g}^{(2)}(0)$ given by
\begin{align}\label{11a}
\mathrm{g}^{(2)}(0)=\frac{Tr[a^\dagger a^\dagger a a\rho]}{Tr[a^\dagger a\rho]^2}.
\end{align}
In the remaining portion of this section, we validate the conformity of our analytical solutions for the optimal UPB effect with the exact numerical results for both cavity-driven and atom-driven cases.
\subsection{\label{sec:level4a}Cavity-driven case}
\begin{figure}[t!]
    \centering
    \includegraphics[width=0.45\textwidth]{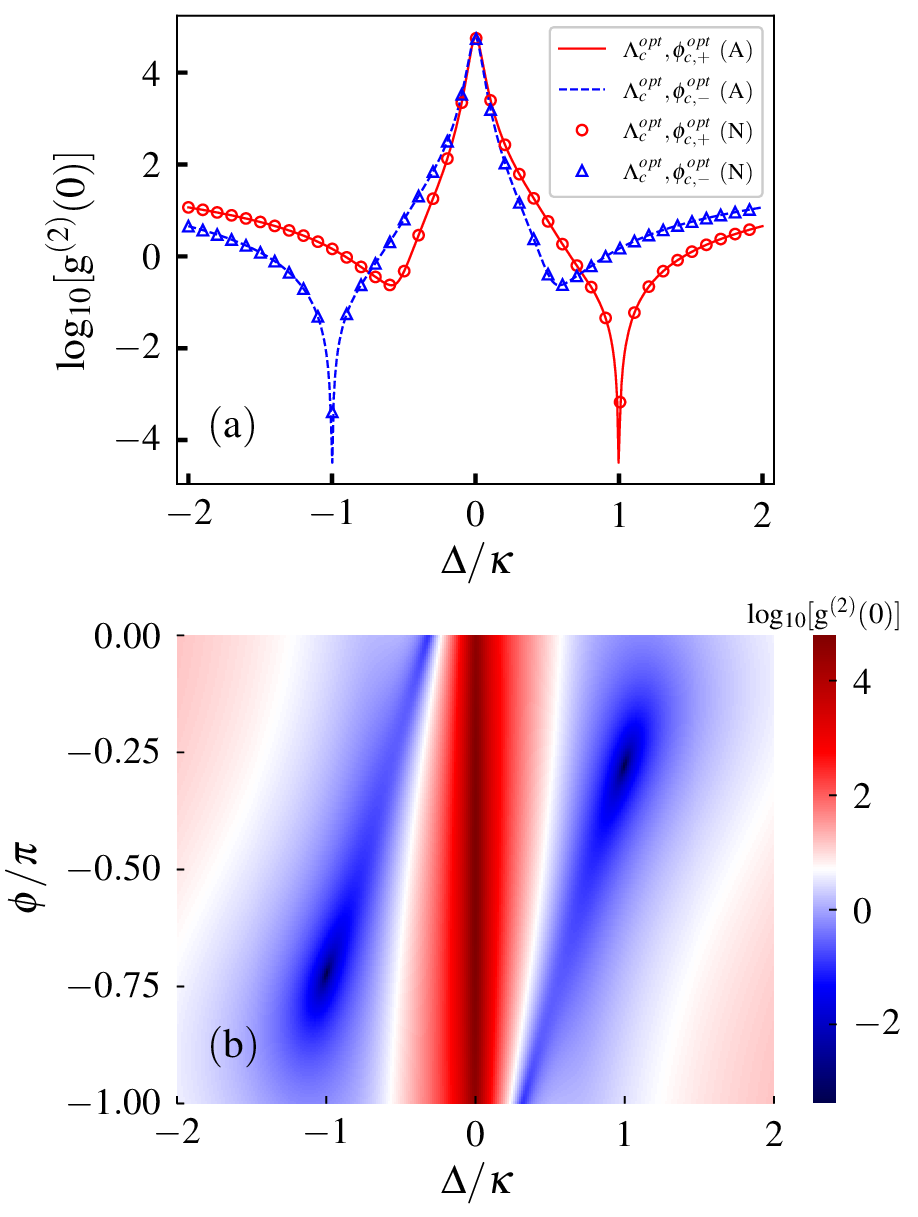}
    \caption{(Color online) The logarithm of the equal-time second-order correlation function $\mathrm{g}^{(2)}(0)$ for the cavity-driven case plotted (a) versus the normalized drive-detuning $\Delta/\kappa$, and (b) in the parametric space of the normalized drive-detuning $\Delta/\kappa$ and the parametric pump phase $\phi$. The analytical (A) (numerical (N)) results of $\mathrm{g}^{(2)}(0)$ in (a) are shown with the solid-red (circle) and dashed-blue (triangle) lines (markers) under the optimal UPB conditions for the drive-detunings $\Delta/\kappa=1$ and $\Delta/\kappa=-1$, respectively. Here, we choose the parameters $\Lambda_{c,\pm}^{opt}/\kappa=\Lambda_{c}^{opt}/\kappa=4.239\times 10^{-5}$, $\phi_{c,+}^{opt}/\pi=-0.277$, and $\phi_{c,-}^{opt}/\pi=-0.723$ in (a), $\Lambda_{c,\pm}^{opt}/\kappa=\Lambda_{c}^{opt}/\kappa=4.239\times 10^{-5}$ in (b), and $\chi/\kappa=1/\sqrt{2}$, $\gamma/\kappa=0.1$, $\Omega/\kappa=0.005$ in both panels.}
    \label{fig:fig3}
\end{figure}
\noindent A simple set of parameters that can be realized experimentally in superconducting quantum circuits is used for the numerical simulations \cite{PhysRevA.92.033806, PhysRevLett.120.083602}. We show in Fig. \ref{fig:fig3} (a) that the minimum numerical equal-time second-order quantum correlation functions can be realized at drive-detunings $\Delta=\pm\kappa$ with the optimal parameters for $\Delta=\pm\kappa$ calculated from Eq. \ref{8} and given in the caption of Fig. \ref{fig:fig3}. The density plot in Fig. \ref{fig:fig3} (b) also suggests that the optimal parameters $\Lambda_{c,\pm}^{opt}$ and $\phi_{c,\pm}^{opt}$ of the parametric pump field results in optimum UPB effect at $\Delta=\pm\kappa$ as predicted by our analytical calculations in Sec. \ref{sec:level3a}. The minimum value of the second-order quantum correlation function $\mathrm{g}^{(2)}(0)$ of the order of $10^{-3}$ is obtained under the optimal UPB conditions as shown in Fig. \ref{fig:fig3} (a) and (b). The destructive quantum interference between the paths $|g,0\rangle\xrightarrow[]{\sqrt{2}\Lambda e^{-i\phi}}|g,2\rangle$, $|g,0\rangle\overset{\Omega}\longrightarrow|g,1\rangle\overset{\sqrt{2}\Omega}\longrightarrow|g,2\rangle$, $|g,0\rangle\overset{\Omega}\longrightarrow|g,1\rangle\overset{\chi}\longleftrightarrow|e,0\rangle\overset{\Omega}\longrightarrow|e,1\rangle\overset{\sqrt{2}\chi}\longleftrightarrow|g,2\rangle$, and other possible two-photon pathways shown in Fig. \ref{fig:fig2} (a) leads to UPB in the cavity-driven system. The optimal parameters for any arbitrary non-zero and zero detuning $\Delta$ can be obtained from Eq. \ref{8}. In the following section, we study the UPB effect in a parametrically pumped, atom-driven cavity QED system, which has not been studied to the best of our knowledge.
\subsection{\label{sec:level4b}Atom-driven case}
\begin{figure}[t!]
    \centering
    \includegraphics[width=0.45\textwidth]{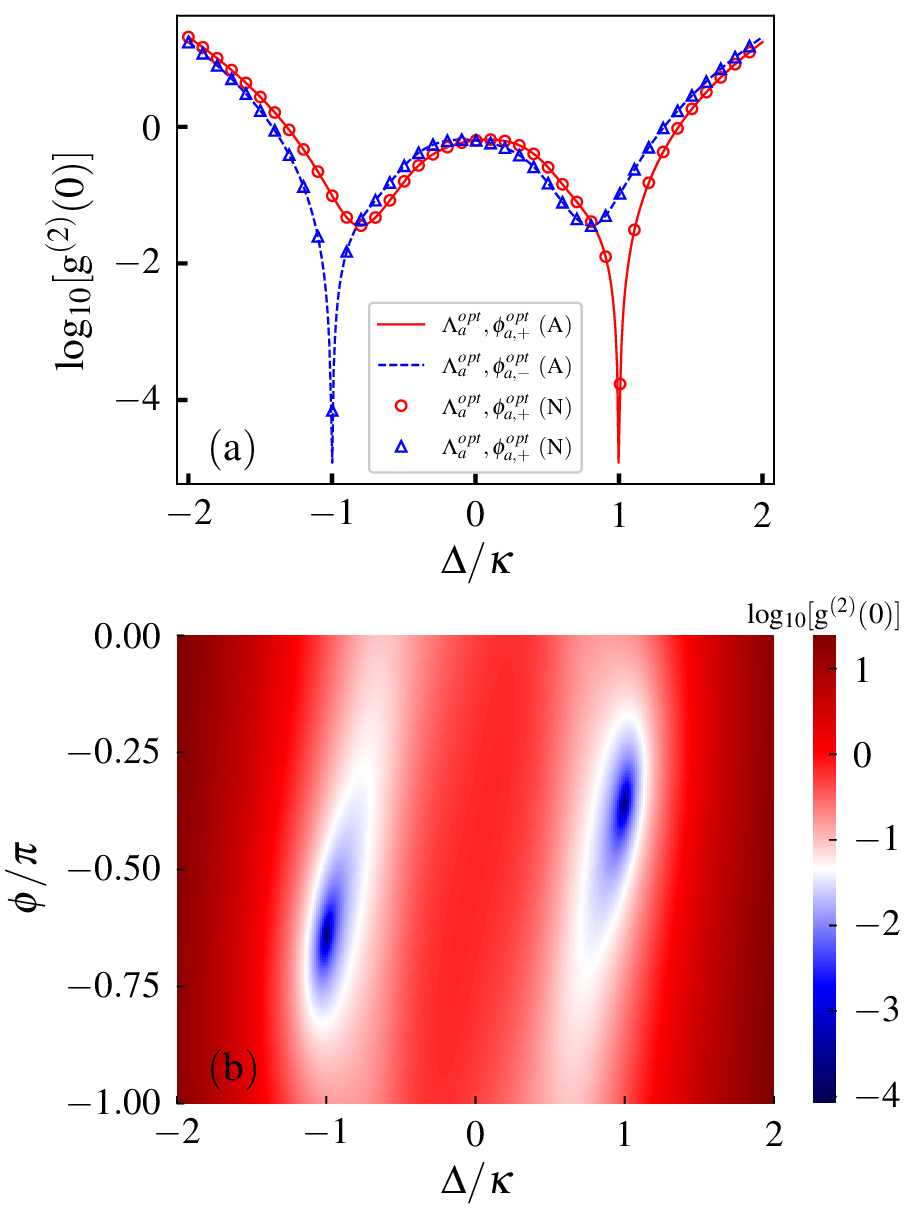}
    \caption{(Color online) The logarithm of the equal-time second-order correlation function $\mathrm{g}^{(2)}(0)$ for the atom-driven case plotted (a) versus the normalized drive-detuning $\Delta/\kappa$, and (b) in the parametric space of the normalized drive-detuning $\Delta/\kappa$ and the parametric pump phase $\phi$. The analytical (A) (numerical (N)) results of $\mathrm{g}^{(2)}(0)$  in (a) are shown with the solid-red (circle) and dashed-blue (triangle) lines (markers) under the optimal UPB conditions for the drive-detunings $\Delta/\kappa=1$ and $\Delta/\kappa=-1$, respectively. Here, we choose the parameters $\Lambda_{a,\pm}^{opt}/\kappa=\Lambda_{a}^{opt}/\kappa=8.292\times 10^{-6}$, $\phi_{a,+}^{opt}/\pi=-0.359$, and $\phi_{a,-}^{opt}/\pi=-0.641$ in (a), $\Lambda_{a,\pm}^{opt}/\kappa=\Lambda_{a}^{opt}/\kappa=8.292\times 10^{-6}$ in (b), and all the other parameters are same as in \ref{fig:fig3}..}
    \label{fig:fig4}
\end{figure}
\noindent The numerical second-order correlation function for the atom-driven case is discussed in this section to verify the analytical analysis in Sec. \ref{sec:level3b}. We obtain different optimal parameters for the UPB effect by considering the atom-driven model system for the exact parameters used for simulations in the cavity-driven case. In Fig. \ref{fig:fig4} (a), the analytical and numerical second-order quantum correlation functions with the optimal parameters $\Lambda_{a,\pm}^{opt}$, and $\phi_{a,\pm}^{opt}$ for $\Delta=\pm\kappa$ in the atom-driven case are plotted. The numerical results conform with the analytical results for the optimal parameters derived from Eq. \ref{11}. The density plot in Fig. \ref{fig:fig4} (b) reiterates the observation as we obtain a minimum second-order quantum correlation function $\mathrm{g}^{(2)}(0)$ of the order of $10^{-4}$ at drive-detunings $\Delta/\kappa=\pm1$ for the optimal parameters $\Lambda_{a,\pm}^{opt}$ and $\phi_{a,\pm}^{opt}$. We note that one can achieve stronger photon antibunching with the optimal parameters for the atom-driven case by using the same set of other system parameters used for the cavity-driven case. The quantum destructive interference between $|g,0\rangle\xrightarrow[]{\sqrt{2}\Lambda e^{-i\phi}}|g,2\rangle$, $|g,0\rangle\overset{\Omega}\longrightarrow|e,0\rangle\overset{\chi}\longleftrightarrow|g,1\rangle\overset{\Omega}\longrightarrow|e,1\rangle\overset{\sqrt{2}\chi}\longleftrightarrow|g,2\rangle$, and other possible two-photon pathways shown in Fig. \ref{fig:fig2} (b) leads to UPB in the atom-driven system. In the rest of this section, we discuss the advantages of the atom-driven system for the UPB effect.

\begin{figure}[t!]
    \centering
    \includegraphics[width=0.4\textwidth]{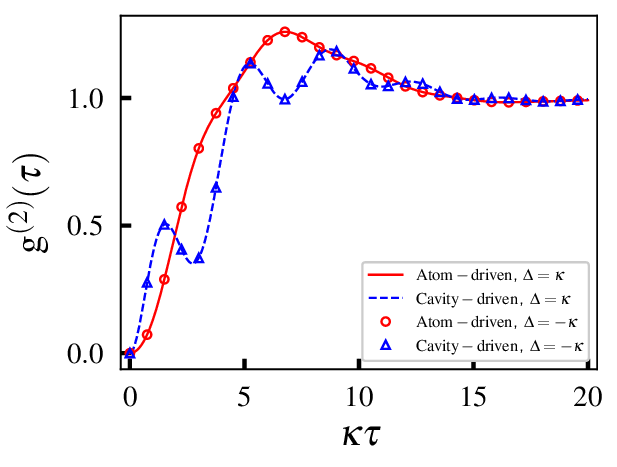}
    \caption{(Color online) The two-time second-order correlation function $\mathrm{g}^{(2)}(\tau)$ versus the normalized time-delay $\kappa\tau$ in the cavity-driven and atom-driven cases for $\Delta/\kappa=\pm 1$ under the optimal UPB conditions with the parameters used in Fig. \ref{fig:fig3} and Fig. \ref{fig:fig4}. The analytical (numerical) results of $\mathrm{g}^{(2)}{(\tau)}$ with $\Delta=\kappa$ ($\Delta=-\kappa$) are shown with the solid-red line (circle markers) and dashed-blue line (triangle markers) for the cavity-driven and atom-driven cases, respectively. }
    \label{fig:fig5}
\end{figure}

The two-time correlation function $\mathrm{g}^{(2)}(\tau)$ can delineate the antibunching of
photons in our model system under the optimal UPB conditions. The two-time second-order correlation function is defined as
\begin{align}\label{12}
\mathrm{g}^{(2)}(\tau)=\frac{\langle a^\dagger(t) a^\dagger(t+\tau) a(t+\tau) a(t)\rangle}{\langle a^\dagger(t) a(t)\rangle^2}.
\end{align}
It is proportional to the probability of detecting two photons at a time delay $\tau$ at the same position in the steady state \cite{10.1093/oso/9780198501770.002.0001}. The necessary condition for the photon antibunching is $\mathrm{g}^{(2)}(\tau)>\mathrm{g}^{(2)}(0)$. Fig. \ref{fig:fig5} depicts the two-time correlation function $\mathrm{g}^{(2)}(\tau)$ for both cavity-driven and atom-driven cases under the optimal UPB conditions for $\Delta=\pm\kappa$. Our numerical result substantiates that one can achieve strong photon antibunching with sub-Poissonian distribution for the optimal parameters with $\mathrm{g}^{(2)}(\tau)>\mathrm{g}^{(2)}(0)$, and $\mathrm{g}^{(2)}(\infty)\rightarrow1$. We also observe more oscillation in $\mathrm{g}^{(2)}(\tau)$ for the cavity-driven case as compared with the atom-driven case. This result is attributed to the difference in the magnitude of optimal parametric gain $\Lambda^{opt}$ with respect to the decay rates $\kappa$, $\gamma$ in the cavity-driven and atom-driven cases \cite{Gerry_Knight_2004}. Higher parametric gain results in higher oscillatory behaviour of the two-time second-order correlation function. 

\begin{figure*}[t!]
    \centering
    \includegraphics[width=0.8\textwidth]{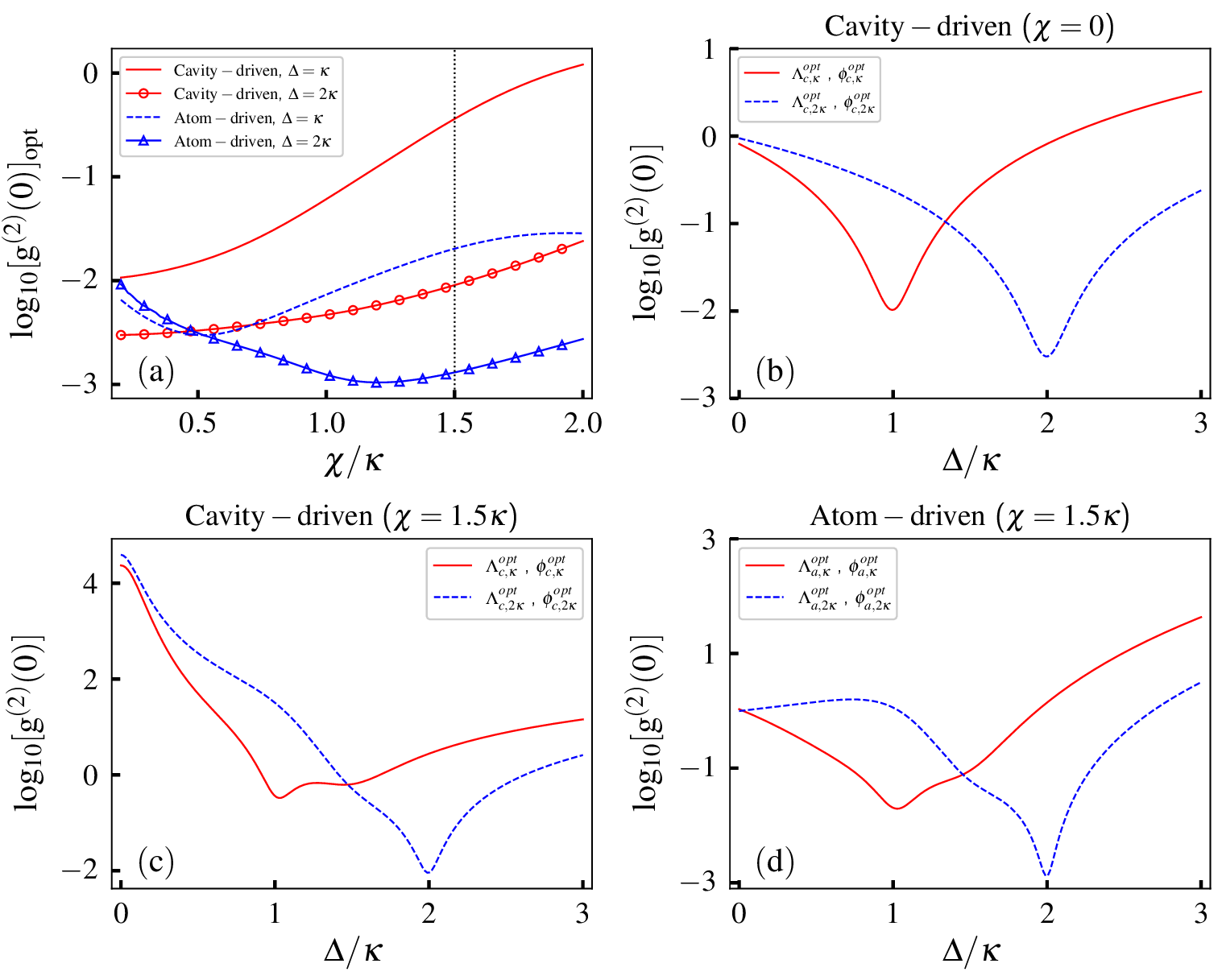}
    \caption{(Color online) The logarithm of the equal-time second-order correlation function $\mathrm{g}^{(2)}(0)$ is plotted versus (a) the normalized cavity-atom coupling strength $\chi/\kappa$ under the optimal UPB conditions for the cavity-driven and atom-driven cases for the drive-detunings $\Delta/\kappa=1$ and $\Delta/\kappa=2$, (b) the normalized drive-detuning under the optimal UPB conditions for $\Delta/\kappa=1,2,$ in the cavity-driven case with the parameters $\Lambda_{c,\kappa}^{opt}/\kappa=2.373\times 10^{-3}$, $\Lambda_{c,2\kappa}^{opt}/\kappa=1.996\times 10^{-3}$, $\phi_{c,\kappa}^{opt}/\pi=-0.907$, $\phi_{c,2\kappa}^{opt}/\pi=-0.184$, and $\chi/\kappa=1.5$, (c) the normalized drive-detuning under the optimal UPB conditions for $\Delta/\kappa=1,2,$ in the atom-driven case with the parameters $\Lambda_{a,\kappa}^{opt}/\kappa=1.249\times 10^{-2}$, $\Lambda_{a,2\kappa}^{opt}/\kappa=4.358\times 10^{-4}$, $\phi_{a,\kappa}^{opt}/\pi=-0.956$, $\phi_{a,2\kappa}^{opt}/\pi=-0.224$, and $\chi/\kappa=1.5$. The parameters $\gamma/\kappa=0.1$ and $\Omega/\kappa=0.04$ are used for all the panels.}
    \label{fig:fig6}
\end{figure*}

Comparing the second-order correlation functions for different cavity-atom coupling strengths $\chi$ and drive-detunings $\Delta$ under the optimal UPB conditions can unveil more information about the advantage of the atom-driven case over the cavity-driven case. To this end, we plot the second-order correlation functions $\mathrm{g}^{(2)}(0)$ versus the normalized cavity-atom coupling strength $\chi/\kappa$ in Fig. \ref{fig:fig6} (a) with drive-detunings $\Delta/\kappa=1,2$, under the optimal UPB conditions for the cavity-driven and atom-driven cases. Under the optimal UPB conditions, the optimum $\mathrm{g}^{(2)}(0)$ monotonically increases by increasing $\chi/\kappa$ for the cavity-driven case. For the atom-driven case, the optimum $\mathrm{g}^{(2)}(0)$ first decreases and then increases with the increase in $\chi$ after attaining a minimum value. The minimum $log_{10}[\mathrm{g}^{(2)}(0)]$ value of $-2.526$ ($-2.982$)  for $\Delta=\kappa$ ($\Delta=2\kappa$) is obtained at $\chi/\kappa=0.516$ ($\chi/\kappa=1.204$). The numerical result suggests that stronger photon antibunching can be realized in the atom-driven system compared to the cavity-driven case with $\Delta/\kappa=1$ with any arbitrary cavity-atom coupling strength $\chi$ under the optimal UPB conditions. However, a stronger photon antibunching can be observed in the cavity-driven case with $\chi/\kappa<0.5$ and in the atom-driven case with $\chi/\kappa>0.5$ for $\Delta/\kappa=2$. These results suggest that the interfering pathways in the atom-driven case are more favourable for quantum destructive interference as compared to the cavity-driven case. Upto one-order lower second-order correlation function can be achieved for the atom-driven case than the cavity-driven case with $\chi/\kappa=1.5$ as shown in Fig. \ref{fig:fig6} (a). Increasing the drive-detuning $\Delta$ also results in a lower optimum second-order correlation function. 

Fig. \ref{fig:fig6} (b) shows the logarithm of the second-order correlation function under the optimal UPB condition at $\Delta/\kappa=1,2$ for the cavity-driven case in the absence of cavity-atom coupling ($\chi=0$). The cavity and atom can be considered separate entities without any cavity-atom interaction. There is the possibility of UPB in the cavity-driven case in the absence of cavity-atom interaction. The external classical drive field and the parametric pumping of the cavity establish multiple pathways for quantum destructive interference, which lead to the UPB effect \cite{PhysRevA.96.053827}. The optimal photon blockade for $\Delta/\kappa=1,2$ can be observed in Fig. \ref{fig:fig6} (b) for the optimal parameters obtained by substituting $\chi=0$, and  $\Delta/\kappa=1,2$ in Eq. \ref{8}. However, the UPB effect can not be realized by driving the atom in the absence of cavity-atom coupling $\chi$. This can also be understood from the transition pathways for the atom-driven case in Fig. \ref{fig:fig2} (b), where only one transition path is possible from $|g,0\rangle\rightarrow|g,2\rangle$ via the parametric pumping of the cavity for $\chi=0$. Thus, the quantum destructive interference for UPB is impossible in the atom-driven case without cavity-atom interaction. We note from Fig. \ref{fig:fig6} (b) that one can achieve a second-order correlation function lower than $10^{-2}$ under the UPB condition for $\Delta/\kappa=2$ without the cavity-atom coupling $\chi$. However, Fig. \ref{fig:fig6} (a) suggests that a second-order correlation function of the order of $10^{-3}$ can be achieved with a non-zero cavity-atom coupling $\chi/\kappa\approx1.2$ under the optimal UPB conditions for $\Delta/\kappa=2$ in the atom-driven case. Therefore, non-zero cavity-atom coupling is beneficial for realizing a stronger photon blockade in the atom-driven case. Increasing the cavity-atom coupling interaction $\chi$ is detrimental to the UPB in the cavity-driven case.

The analytical results for optimal parameters produced in Sec. \ref{sec:level3} in the weak cavity-atom coupling limit are shown to be valid for $\chi>\kappa$ as demonstrated in Fig. \ref{fig:fig6} (c), (d). However, the analytical solutions are not valid for $\chi>2\kappa$, which can be verified with the numerical simulations. The logarithm of the second-order correlation functions for the cavity-driven and atom-driven cases for parameters $\chi/\kappa=1.5$ and $\Delta/\kappa=1,2$ under the optimal UPB conditions are illustrated in Fig. \ref{fig:fig6} (c) and (d), respectively. The lowest second-order correlation functions $\chi^{(2)}(0)$ are achieved at $\Delta/\kappa=1,2$ in Fig. \ref{fig:fig6} (c) and (d) by using the analytically calculated optimal parametric gain and phase parameters at the respective detunings for the cavity-driven and atom-driven cases, respectively. It shows the consistency of the numerical results with the analytical calculations for $\chi/\kappa=1.5$. The effect of a further increase in drive-detuning on the UPB effect needs further investigation.
\section{\label{sec:level5}Conclusion}
\noindent We theoretically investigated the optimal conditions for UPB in a parametrically pumped two-level atom-cavity system, with a simultaneous coherent drive applied to the cavity or the atom. The parametric pumping introduces an additional two-photon excitation pathway that results in strong photon antibunching due to quantum destructive interference in cavity- and atom-driven cases. The parametric pump's optimal gain and relative phase difference with respect to the coherent drive field are analytically calculated via the amplitude method for the optimal UPB in the cavity-driven and atom-driven configurations of the proposed model system. We numerically evaluated the second-order quantum correlation function $\mathrm{g}^{(2)}(0)$ with QUTIP \cite{JOHANSSON20131234} to demonstrate UPB with strong photon antibunching. The numerical results of $\mathrm{g}^{(2)}(0)$ confirm the approximate analytical results for both cavity-driven and atom-driven cases. Furthermore, the two-time second-order correlation functions $\mathrm{g}^{(2)}(\tau)$ confirmed the antibunching of photons under the optimal UPB conditions.  The proposed model is experimentally feasible for realizing the UPB effect because tuning the parametric gain and the relative phase difference is more accessible than controlling the drive-detuning and cavity-atom coupling. We demonstrated that coherently driving the atom results in up to one order ($\sim 10^{-1}$) reduction in optimal second-order correlation function $\mathrm{g}^{(2)}(0)$ as compared with the cavity-driven system. The non-zero cavity-atom interaction $\chi$ benefits UPB over the zero cavity-atom coupling in the atom-driven case. However, the presence of a non-zero cavity-atom coupling is shown to be detrimental to the UPB effect in the cavity-driven scenario. Setting the cavity-atom coupling strength $\chi$ and drive-detuning $\Delta$ at specific values under the optimal UPB condition for the atom-driven system may result in a lower second-order correlation function. The experimental realization of the proposed model system is well within reach of the current state-of-the-art solid-state quantum devices. 

\appendix
\section{\label{app:level16} Derivation of optimal parameters}
\noindent To solve the coupled equations given in Eq. \ref{5}, and Eq. \ref{9} in the steady state, we consider the perturbative approximations $|C_{g,0}|\gg|C_{g,1}|,|C_{e,0}|\gg|C_{g,2}|,|C_{e,1}|$, and $|C_{g,0}|\approx 1$. Therefore, the steady-state equations of motion (\textbf{EOM}) for single-photon states in Eq.  \ref{5} can be written as
\begin{align}\label{A1}
 \Delta_c^\prime C_{g,1}+\chi C_{e,0}+\Omega & = 0 \nonumber\\
 \Delta_a^\prime C_{e,0}+\chi C_{g,1} & = 0
\end{align}
Similarly, one can write the steady-state EOM for the single-photon states in Eq. \ref{9} as
\begin{align}\label{A2}
 \Delta_c^\prime C_{g,1}+\chi C_{e,0} & = 0\nonumber\\
 \Delta_a^\prime C_{e,0}+\chi C_{g,1}+\Omega & =0
\end{align}
Solving Eq. \ref{A1} and Eq. \ref{A2} give the probability steady-state amplitudes $C_{g,1}$, $C_{e,0}$, in the cavity-driven and atom driven cases, respectively. The probability amplitudes of two-photon states $C_{e,1}$ and $C_{g,2}$ can be calculated by substituting the steady-state values of $C_{g,1}$, $C_{e,0}$ in the steady-state EOM of two-photon states.

\bibliographystyle{apsrev4-2}
\bibliography{ref}
\end{document}